\documentstyle[preprint,aps]{revtex}
\begin{document}
\tighten
\draft
\preprint{PSU/TH/176; hep-ph/9611399}
\title{Recoil Corrections of Order $(Z\alpha)^6(m/M)m$ to the Hydrogen 
Energy Levels Revisited} 
\author {Michael I. Eides \thanks{E-mail address:  
eides@phys.psu.edu, eides@lnpi.spb.su}}
\address{ Department of Physics, Pennsylvania 
State University, 
University Park, PA 16802, USA\thanks{Temporary address.}\\ 
and
Petersburg Nuclear Physics Institute,
Gatchina, St.Petersburg 188350, Russia\thanks{Permanent address.}}
\author{Howard Grotch\thanks{E-mail address: h1g@psuvm.psu.edu}}
\address{Department of Physics, Pennsylvania State University,
University Park, PA 16802, USA}
\date{November, 1996}

\maketitle
\begin{abstract}
The recoil correction of order $(Z\alpha)^6(m/M)m$ to the hydrogen energy 
levels is recalculated and a discrepancy existing in the literature on this 
correction for the $1S$ energy level, is resolved. An analytic expression 
for the correction to the $S$-levels with arbitrary principal quantum number 
is obtained.  
\end{abstract} 

\section{Introduction}

The calculation of the recoil corrections of order $(Z\alpha)^6(m/M)m$ to 
the hydrogen energy levels has a long history 
Refs.\cite{gy,eg,dgo,dge,kmy,fkmy}.  After initial disagreements consensus 
was achieved in Ref.\cite{pg}, where one and the same result was 
obtained in two apparently different frameworks.  The first, more 
traditional approach, used earlier in Refs.\cite{eg,dgo,dge}, starts with an 
effective Dirac equation in the external field.  Corrections to the Dirac 
energy levels are calculated with the help of a systematic diagrammatic 
procedure. The other logically independent calculational framework, also 
used in Ref.\cite{pg}, starts with an exact expression for all recoil 
corrections of the first order in the mass ratio of the light and heavy 
particles $m/M$. This remarkable expression, which is exact in $Z\alpha$, 
was first discovered by M. A. Braun \cite{braun}, and rederived later in 
different ways in a number of papers Refs.\cite{shab,yelkh,pg}. 

The agreement on the $(Z\alpha)^6(m/M)m$ contribution achieved in 
\cite{pg} seemed to put an end to all problems connected with this 
correction.  However, it was claimed in a recent work \cite{elkh}, that 
the result of \cite{pg} is in error. The discrepancy between the results 
of Refs.\cite{pg,elkh} is confusing since the calculation in \cite{elkh} is 
performed in the same framework as the one employed in \cite{pg}, namely it 
is based on a particularly nice form of the Braun formula obtained by the 
author earlier \cite{yelkh}, 

\begin{equation}                      \label{braun}
\Delta E_{rec}=-\frac{1}{M}Re\int\frac{d\omega}{2\pi i}
<{n}|({\bf p}-{\bf\hat D}(\omega))G(E+\omega)({\bf 
p}-{\bf\hat D}(\omega))|n>, 
\end{equation}

where summation over all intermediate states is understood, $G(E+\omega)$ is 
the Coulomb-Green function in the Coulomb gauge, which in the momentum space 
has the form

\begin{equation}         
{\bf\hat D}(\omega,k)=-4\pi Z\alpha(\mbox{\boldmath$\alpha$}-\frac{{\bf 
k}(\mbox{\boldmath$\alpha k$})}{{\bf k}^2})\frac{1}{\omega^2-{\bf k}^2+i0}
\equiv -4\pi Z\alpha\frac{\mbox{\boldmath$\alpha_k$}}{\omega^2-{\bf k}^2+i0},
\end{equation}

and

\begin{equation} 
\alpha_i=\gamma^0\gamma^i.
\end{equation}

Note that ${\bf\hat D}(\omega,k)$ is nothing more than the transverse photon 
propagator with the source at the proton position, and integration over the 
exchanged photon momentum $\bf k$ is implicit in the expression above. 
Below we will explicitly perform multiplication in the matrix element in 
Eq.(\ref{braun}).  Respective contributions to the energy levels will be 
called Coulomb (corresponds to $\bf pp$), magnetic (corresponds to $\bf 
p\hat D$ and $\bf\hat Dp$), and seagull (corresponds to $\bf\hat D \hat D$).

It is the aim of this paper to resolve the above noted discrepancy on the 
recoil correction of order $(Z\alpha)^6(m/M)m$ to the $1S$ energy level, and 
also to obtain this correction for the $S$-levels with arbitrary principal 
quantum number (it was earlier calculated only for $n=1,2$ \cite{pg}).

\section{Two Approaches to the Braun Formula}

Calculation of the recoil contribution of order $(Z\alpha)^6$ generated by 
the Braun formula was performed in \cite{pg} in a most straightforward 
way since separation of the high- and low-frequency contributions was made 
in the framework of the $\epsilon$-method developed by one of the authors 
earlier \cite{pach}. Hence, not only were contributions of order 
$(Z\alpha)^6(m/M)m$ obtained in Ref.\cite{pg}, but also linear in $m/M$ 
parts of recoil corrections of orders $(Z\alpha)^4$ and $(Z\alpha)^5$ 
(ref.\cite{salp}) were reproduced for the $1S$-state. Note that the Braun 
formula, despite its obvious advantages, in its present form sums only 
contributions linear in the mass ratio. Hence, old methods are more adequate 
for obtaining the proper mass dependence of the contributions of orders 
$(Z\alpha)^4$ and $(Z\alpha)^5$, which were worked out in Ref.\cite{gy}. 
Calculations in Ref.\cite{pg} turned out to be rather lengthy and tedious 
just because all corrections of previous orders in $Z\alpha$ were 
reproduced.

The most significant feature of the recoil corrections of order 
$(Z\alpha)^6$, which made the whole approach of Ref.\cite{elkh} possible, is 
connected with the absence of {\it logarithmic} recoil corrections of 
this order, as was proved in \cite{fkmy}. Unlike \cite{pg}, the 
calculations in \cite{elkh} are organized in such a way that one explicitly 
makes approximations inadequate for calculation of the contributions of 
the previous orders in $Z\alpha$, significantly simplifying calculation 
of the correction of order $(Z\alpha)^6$. Due to absence of the logarithmic 
contributions of order $(Z\alpha)^6$, infrared divergences connected with 
the crude approximations unadequate for calculation of the contributions of 
the previous orders would be powerlike and can be safely thrown away. Next, 
absence of logarithmic corrections of order $(Z\alpha)^6$ means that it is 
not necessary to worry too much about matching the low- and high-frequency 
(long- and short-distance in terms of Ref.\cite{elkh}) contributions, since 
each region will produce only nonlogarithmic contributions and correction 
terms would be suppressed as powers of the separation parameter. We would 
like to emphasize once more that this approach would be doomed if the 
logarithmic divergences were present, since in such a case one could not 
hope to calculate an additive constant to the log, since the exact value of 
the integration cutoff would not be known.

We are going to perform below calculation of the recoil contribution of 
order $(Z\alpha)^6$ in the framework of Ref.\cite{elkh}, and to discover the 
source of discrepancy between the results of Ref.\cite{pg} and 
Ref.\cite{elkh}.  In order to really implement such program we need to have 
a regular method to qualify all terms which will be thrown away. To this end 
we will use a slight generalization of the ordinary approach to calculation 
of the leading order contribution to the Lamb shift. 

It may be proved that all corrections of order $(Z\alpha)^6(m/M)m$ are 
generated by the exchange of photons with momenta larger than 
$m(Z\alpha)^2$, so we will consider below only this integration region. In 
the spirit of the common approach to the Lamb shift calculations we will 
split the integration region over the exchanged photon momenta (and when 
necessary over frequencies) with the help of an auxiliary parameter $\sigma$ 
which satisfies the conditions

\begin{equation}
mZ\alpha\ll\sigma\ll m,
\end{equation}

and we will call the photons with momenta smaller than $\sigma$ 
low-frequency (or long-distance) photons, and the photons with momenta 
larger than $\sigma$ will be called high-frequency (or short-distance) 
photons. Considering low-frequency photons we may expand over the ratio 
$k/m$ since for such photons $k/m\leq\sigma/m\ll 1$\footnote{Note that the 
apparent linear divergences in this region of the form $\sigma/m$ are 
really parametrically small.}.  On the other hand, for the high-frequency 
photons $mZ\alpha/k\leq mZ\alpha/\sigma\ll1$, and we may expand over this 
parameter. Note that for momenta of order $\sigma$ both expansions are 
valid simultaneously, and, hence, we may match the expansions and get rid of 
the auxiliary parameter $\sigma$. However, the problem under consideration 
is in a sense even simpler than calculation of the leading order 
contribution to the Lamb shift, and due to absence of the logarithmic 
contributions of order $(Z\alpha)^6(m/M)m$, precise matching of the high- 
and low-frequency contributions is unnecessary. Below we will consider 
calculation only of the low-frequency ($mZ\alpha<k<\sigma$) contribution 
to the energy shift, since for the high frequency contribution the results 
of Ref.\cite{pg} and Ref.\cite{elkh} nicely coincide.

\section{Main Recoil Contribution}

With the help of the Braun formula one may easily obtain an expression for 
the leading recoil correction which is linear in the mass ratio and which 
includes all terms of order $(Z\alpha)^4$ and lower (see Ref.\cite{shab}). 
To this end we rewrite the Coulomb contribution in Eq.(\ref{braun}) in the 
form

\begin{equation}
\Delta E_{Coul}=
\frac{1}{2M}<{n}|{\bf p}^2|n>-\frac{1}{M}<{n}|{\bf p}\Lambda^-{\bf p}]|n>
\end{equation}
\[
\equiv \Delta E_{c1}+\Delta E_{c2}.
\]

We also extract the nonretarded Breit part from the magnetic contribution in 
Eq.(\ref{braun} )

\begin{equation}
\Delta E_{magn}=\Delta E_{Br}+\Delta E_{magn,r},
\end{equation}

where

\begin{equation}
\Delta E_{Br}=
-\frac{1}{2M}
<{n}|{\bf p}{\bf\hat D}(0,k)
+{\bf\hat D}(0,k){\bf p}|n>,
\end{equation}

and

\begin{equation}
\Delta E_{magn,r}=
-\frac{1}{M}\int\frac{d\omega}{2\pi i}
<{n}|[V,{\bf p}]G(E+\omega){\bf\hat D}(\omega,k)
\end{equation}
\[
-{\bf\hat D}(\omega,k)G(E+\omega)[V,{\bf p}]|n> 
\frac{1}{\omega+i0},
\]

where $V$ is the Coulomb potential ($V=-Z\alpha/r$).

Now it is not difficult to check with the help of the virial relations (see, 
e.g., Ref.\cite{ee}), that the sum of the main part of the Coulomb term and 
of the Breit contribution acquires a very nice form

\begin{equation}                  \label{gys}
\Delta E_{c1}+\Delta E_{Br}=\frac{m^2-E^2}{2M},
\end{equation}

where $E$ is the value of the energy given by the Dirac equation. As we will 
see below, all other recoil contributions to the energy level start at 
least with the term of order $(Z\alpha)^5$, and, hence, the formula above 
correctly describes all contributions of order $(Z\alpha)^4$ and lower. 
However, this formula describes only contributions linear in the mass ratio.  
A more precise expression which takes into account corrections of  
higher order in $m/M$, was obtained in Ref.\cite{gy}. 

It is easy to see that the expression in Eq.(\ref{gys}) also contains 
the correction of order $(Z\alpha)^6$, which for the $nS$-states has the form

\begin{equation}    \label{gy}
\Delta 
E_{GY}=(\frac{1}{8}+\frac{3}{8n}-\frac{1}{n^2}+\frac{1}{2n^3})
\frac{(Z\alpha)^6}{n^3}\frac{m}{M}m.  
\end{equation}

This contribution was originally obtained in Ref.\cite{gy}.

The remaining part of the Coulomb contribution has the form

\begin{equation}
\Delta E_{c2}=-\frac{1}{M}<n|{\bf p}\Lambda_-{\bf p}|n>.
\end{equation}

Let us check that this term leads to corrections of higher order than 
$(Z\alpha)^6$ when the intermediate momenta are of the atomic scale. We want 
to exploit the large (of order $2m$) value of the energy gap between 
positive and negative states in comparison with the typical energy 
splittings (of order $m(Z\alpha)^2$) in the positive energy spectrum. First, 
let us note that

\begin{equation}
<n|[{\bf p},V]\Lambda_-[{\bf p},V]|n>=<n|[{\bf p},H-E]\Lambda_-[{\bf 
p},H-E]|n>
\end{equation}
\[
= -<n|{\bf p}\sum_-|m><m|(E_n-E_m)^2{\bf p}|n>.  
\]

However, $(E_n-E_m)^2>4m^2(1-c\alpha^2)$, and, hence, 

\begin{equation}
|<n|[{\bf p},V]\Lambda_-[{\bf p},V]|n>|=|<n|{\bf 
p}\sum_-|m><m|(E_n-E_m)^2{\bf p}|n>| 
\end{equation}
\[
\geq |<n|{\bf p}\Lambda_-{\bf p}|n>|4m^2(1-c\alpha^2).
\]

Then 

\begin{equation}
|<n|{\bf p}\Lambda_-{\bf p}|n>|\leq\frac{1}{4m^2(1-c\alpha^2)}
|<n|[{\bf p},V]\Lambda_-[{\bf p},V]|n>|.
\end{equation}

We know that at the atomic scale the Coulomb potential is of order 
$(Z\alpha)^2$, the momentum operators are of order $Z\alpha$, and, hence, we 
explicitly have the factor $(Z\alpha)^6$. Note that this approach would not 
work if we had a projector on the positive energy states. In such a case the 
energy differences would be of order $(Z\alpha)^2$ themselves and we would 
not get any suppression, since the factors $(Z\alpha)^2$ would cancel in the 
numerator and denominator.

Returning to our case, it is easy to realize that the projector on the 
negative energy states leads to additional suppression in the 
nonrelativistic limit, and, hence, the term under consideration does not 
produce any contribution of order $(Z\alpha)^6$ at the atomic scale.

There is complete agreement between the results of Ref.\cite{pg} and 
Ref.\cite{elkh} for the corrections discussed in this section.

\section{Seagull Contribution}

Following Ref.\cite{elkh} let us again start with the Braun expression 
Eq.(\ref{braun}) for the seagull contribution and perform the integration by 
closing the contour each time around one of the transverse photon poles

\begin{equation}  \label{seagull}
\Delta E_{s}=-\frac{1}{M}\int\frac{d\omega}{2\pi i}
<{n}|{\bf\hat D}(\omega,k)G(E+\omega){\bf\hat D}(\omega,k)|n>.
\end{equation}

Substituting the pole representation for the Coulomb Green function 
we obtain in accordance with Ref.\cite{elkh}

\begin{equation}      \label{seagullpos}
\Delta E_{s}=
\frac{(Z\alpha)^2}{2M}
<{n}|\frac{4\pi\mbox{\boldmath 
$\alpha_{k'}$}}{k'}[
\sum_+\frac{|m><m|}{(E-k'-E_m)(E-k-E_m)}(1+\frac{E_m-E}{k'+k})
\end{equation} 
\[ 
-\sum_-\frac{|m><m|}{(E+k'-E_m)(E+k-E_m)}(1-\frac{E_m-E}{k'+k})]
\frac{4\pi\mbox{\boldmath $\alpha_k$}}{k}|n>. 
\]

Let us consider positive- and negative-energy parts of this expression 
separately.

We may expand the positive energy part in $(E-E_m)/k$ and $(E-E_m)/k$, 
taking into account that in the low-frequency integration region 
$mZ\alpha<k<\sigma$. In the first order of this expansion we get

\begin{equation}
\Delta E^+_{s}=\frac{(Z\alpha)^2}{2M}
<{n}|\frac{4\pi\mbox{\boldmath $\alpha_{k'}$}}{k'^2}
\Lambda_+\frac{4\pi\mbox{\boldmath $\alpha_k$}}{k^2}|n>. 
\end{equation}

Calculation of this contribution will be considered below.
Let us turn to the negative-energy contribution. 
Energy differences are large for the negative energy contribution 
($|E-E_m|\approx 2m(1-c\alpha^2)$), so we expand the negative 
energy term in $k/(E-E_m)$

\begin{equation}
\sum_-\frac{|m><m|}{(E+k'-E_m)(E+k-E_m)}(1-\frac{E_m-E}{k'+k})
\end{equation}
\[
=\sum_-\frac{|m><m|}{E-E_m}[\frac{1}{k+k'}+\frac{(k+k')^2}{2(E-E_m)^3}].
\]

In accordance with Ref.\cite{elkh} the terms linear in $k/2m$  cancel, 
and the negative energy contribution acquires the form

\begin{equation}
\Delta E^-_{s}
=-\frac{(Z\alpha)^2}{4mM(1+c\alpha^2)}
<{n}|\frac{4\pi\mbox{\boldmath $\alpha_{k'}$}}{k'}
\Lambda_-[\frac{1}{k+k'}+\frac{(k+k')^2}
{2[2m(1+c\alpha^2)]^3}]
\frac{4\pi\mbox{\boldmath $\alpha_k$}}{k}|n>. 
\end{equation}

It may be shown (compare below consideration of the negative energy 
contribution in the case of the one transverse exchange) that the first term 
produces the contribution of order $(Z\alpha)^5$ while the second is of 
order $(Z\alpha)^7$. Only terms linear in $k$, $k'$ are capable of
producing contributions of order $(Z\alpha)^6$, but these terms cancel each 
other, as we have just seen.

Let us now return to the positive energy contribution. The idea of 
Ref.\cite{elkh} is to consider matrix elements and to calculate them in the 
nonrelativistic approximation, which produces the leading low-frequency 
contribution. All matrix elements under consideration have common structure. 
In general they are the products of matrix elements of $\gamma$-matrices in 
the momentum space.  Each such matrix element in the nonrelativistic limit 
may easily be reduced to an explicit function of momenta and 
$\sigma$-matrices, then transformed into coordinate space and calculated 
between Coulomb-Schrodinger wave functions.

We have performed an explicit calculation along these lines and obtained in 
complete accord with Ref.\cite{elkh} 

\begin{equation} \label{seagullop}
\Delta E^+_{s}=\frac{(Z\alpha)^2}{4m^2M}<n|2{\bf 
p}\frac{1}{r^2}{\bf p}+\frac{1}{r^4} -\frac{3{\bf l}^2+2\mbox{\boldmath 
$\sigma l$}}{2r^4}|n>.
\end{equation}

This expression is singular at the origin. This singularity 
produces  linear and logarithmic ultraviolet divergences in momentum 
space as well as a constant contribution, and hence the contribution 
under consideration cannot be calculated unambiguously in the general case. 
It is necessary to realize at this stage that the initial expression for the 
seagull contribution in Eq.(\ref{seagull}) was defined unambiguously. Even 
separation of the integration region with the help of the auxiliary 
parameter $\sigma$ could not lead to an ultraviolet divergence in the 
low-frequency region since all momentum integrations are cut off from above 
by $\sigma$ and should generate not power divergent but power suppressed 
terms.  It is clear that the apparent divergence is connected with our 
inaccurate calculation of the singularity at large momenta or small 
distances. Hence, we have to return to the initial 
momentum space expression for the positive energy seagull contribution and 
perform all calculations directly in the momentum space. The result of such 
a calculation may be later interpreted as an unambiguous prescription for 
the proper regularization of the coordinate space operators for the 
$S$-states.

Note, that for the non-$S$ states, wave functions vanish at the origin, 
the operators above are well defined on such wave functions, and lead to 
unambiguous results. Of course, any regularization at small distances will 
not influence the value of the non-$S$ matrix elements of the operator in 
Eq.(\ref{seagullop}), and will not influence the agreement between the 
$P$-level energy shift calculated in Ref.\cite{elkh}, and the same shift 
obtained earlier in another framework in Ref.\cite{gkmy}.

\subsection{Accurate Calculation with Momentum Space Cutoff}

Direct calculation of the positive energy seagull contribution  
Eq.(\ref{seagullpos}) in momentum space leads to the following 
expression for the $S$-state contribution

\begin{equation}
\Delta E^+_{s}=
=\frac{(Z\alpha)^2}{m^2M}\int 
\frac{d^3p'}{(2\pi)^3}\frac{d^3p}{(2\pi)^3}\frac{d^3k'}{(2\pi)^3}
\frac{d^3k}{(2\pi)^3}(2\pi)^3\delta({\bf p'-p-k-k'})\frac{8\pi^2}{k'^2k^2}
\end{equation}
\[ 
\psi(p')[-{\bf p'p} +\frac{\bf (k'k)(p'k') (pk)}{k'^2k^2}-\frac{\bf 
k'k}{2}]\psi(p)\equiv \Delta E_{s1}+\Delta E_{s2}+\Delta E_{1/r^4}.
\]

The first two terms in the integrand do not rise too rapidly with $k$ and 
$k'$, and we may unambiguously calculate them using the Fourier transforms 
discussed above. For the first term we have

\begin{equation} 
\Delta E_{s1}=
-\frac{(Z\alpha)^2}{m^2M}\int d^3r
\frac{d^3p'}{(2\pi)^3}\frac{d^3p}{(2\pi)^3}\frac{d^3k'}{(2\pi)^3}
\frac{d^3k}{(2\pi)^3}e^{i\bf r(-p'+p+k+k')})\frac{8\pi^2}{k'^2k^2}
\end{equation}
\[ 
\psi(p'){\bf p'p}\psi(p)
=-\frac{(Z\alpha)^2}{2m^2M}\int d^3r
\frac{d^3p'}{(2\pi)^3}\frac{d^3p}{(2\pi)^3}e^{i\bf r(-p'+p)})
\frac{1}{r^2}
\psi(p'){\bf p'p}\psi(p).
\]

The remaining integration over $p'$ and $p$ simply returns us to the  
coordinate space wave functions, and we may rewrite the expression above in 
the operator notation\footnote{One has to take into account that the 
apparent sign of the expression below changes, since the momenta in the 
exponent have opposite signs.}

\begin{equation} \label{s1}
\Delta E_{s1}=\frac{(Z\alpha)^2}{2m^2M}<n|{\bf p}\frac{1}{r^2}{\bf p}|n>.
\end{equation}

This contribution exactly reproduces the nonsingular operator obtained in 
the previous chapter.

Next we calculate the second contribution in the same manner as above

\begin{equation}
\Delta E_{s2}
=\frac{(Z\alpha)^2}{m^2M}\int d^3r\int 
\frac{d^3p'}{(2\pi)^3}\frac{d^3p}{(2\pi)^3}\frac{d^3k'}{(2\pi)^3}
\frac{d^3k}{(2\pi)^3}
e^{i\bf r(-p'+p+k+k')}
\psi(p')\frac{8\pi^2\bf (k'k)(p'k') (pk)}{k'^4k^4}\psi(p)
\end{equation}
\[ 
=\frac{(Z\alpha)^2}{2m^2M}\int d^3r\int 
\frac{d^3p'}{(2\pi)^3}\frac{d^3p}{(2\pi)^3}
e^{i\bf r(-p'+p)})
\psi(p')\frac{p'_jp_m}{4r^2}(\delta_{ij}-\frac{r_ir_j}{r^2})
(\delta_{im}-\frac{r_ir_m}{r^2})\psi(p)
\]
\[
=\frac{(Z\alpha)^2}{2m^2M}\int d^3r\int 
\frac{d^3p'}{(2\pi)^3}\frac{d^3p}{(2\pi)^3}
e^{i\bf r(-p'+p)})
\psi(p')\frac{1}{4r^2}({\bf p'p}-\frac{\bf(p'r)(pr)}{r^2})\psi(p).
\]

Now we use the formula 

\begin{equation}
{\bf (rp')(rp)=-[r\times p'][r\times p]+r^2(p'p)},
\end{equation}

and omit the terms with the vector product since we are considering only 
$S$-states now. Then we obtain

\begin{equation}
\Delta E_{s2}=0.
\end{equation}

Next we have to calculate the third contribution, which corresponds to the 
$1/r^4$ term in the naive result above in Eq.(\ref{seagullop}). This time we 
cannot use Fourier transformations over exchanged momenta for calculation of 
this integral, since this leads to a singular expression in coordinate 
space.  So we first perform the safe Fourier transformations over the wave 
function momenta, and then directly evaluate the exchanged momenta 
integrals, taking into account that they are cut from above by $\sigma\ll m$,

\begin{equation}
\Delta E^{1/r^4}_{s}
=-\frac{(Z\alpha)^2}{m^2M}
\int \frac{d^3k'}{(2\pi)^3}
\frac{d^3k}{(2\pi)^3}\frac{4\pi^2({\bf k'k})}{k'^2k^2}
<n({\bf r})|e^{i{\bf (k+k')r}}|n({\bf r})>.
\end{equation}

In order to preserve the transparency of the presentation we will perform 
the calculation only for $n=1$ here. The general case of arbitrary principal 
quantum number will be considered at the end of the paper. We 
substitute explicit expressions for the $1S$-wave functions in the 
formula above, and do the coordinate-space integral 

\begin{equation}
\Delta E^{1/r^4}_{s}=
-\frac{(Z\alpha)^2}{m^2M}|\psi(0)|^2
\int\frac{d^3k'}{(2\pi)^3}
\frac{d^3k}{(2\pi)^3}\frac{4\pi^2({\bf k'k})}{k'^2k^2}\int d^3r
e^{i{\bf (k+k')r}}e^{-2\gamma r}
\end{equation}
\[
=-\frac{64\pi^3(Z\alpha)^2}{m^2M}\gamma|\psi(0)|^2
\int \frac{d^3k'}{(2\pi)^3}
\frac{d^3k}{(2\pi)^3}\frac{{\bf k'k}}{k'^2k^2[({\bf 
k+k'})^2+(2\gamma)^2]^2},  
\]

where $\gamma=mZ\alpha$.

Symmetrical integrals over the exchanged momenta are cut from above by the 
parameter $\sigma$. However, first integration, say over $\bf k'$, is 
convergent at high momenta and the cutoff may be safely ignored

\begin{equation}           \label{s3}
\Delta E^{1/r^4}_{s}=-\frac{16\pi(Z\alpha)^2}{m^2M}\gamma|\psi(0)|^2
\int \frac{d^3k}{(2\pi)^3k^2}\int_0^\infty 
dk'\int_{-1}^1dx\frac{k'kx}{[k^2+k'^2+2kk'x+(2\gamma)^2]^2} 
\end{equation} 
\[
=-\frac{8\pi^2(Z\alpha)^2}{m^2M}\gamma|\psi(0)|^2
\int 
\frac{d^3k}{(2\pi)^3k^2}[\frac{arctan\frac{k}{2\gamma}}{k}-\frac{1}{2\gamma}] 
=-\frac{4(Z\alpha)^2}{m^2M}\gamma|\psi(0)|^2
\int_0^\sigma dk[\frac{arctan\frac{k}{2\gamma}}{k}-\frac{1}{2\gamma}] 
\]
\[
=-\frac{(Z\alpha)^2}{m^2M}\gamma|\psi(0)|^2
[2{\pi}\ln\frac{\sigma}{2\gamma} -2\frac{\sigma}{\gamma}].
\]

Nonlogarithmic term of order $(Z\alpha)^5$ in this expression is 
additionally suppressed by the small ratio $\sigma/m$, and may be safely 
ignored.  Thus, we see that the properly regularized operator $1/r^4$ in the 
seagull diagram does not generate a constant contribution.  The logarithmic 
divergence above should cancel with the respective contribution of the 
one-transverse (magnetic) diagram.

\section{Magnetic Contribution}

This time we start with the Braun expression for the one transverse photon 
in Eq.(\ref{braun})

\begin{equation}
\Delta E_{magn}=
\frac{1}{M}Re\int\frac{d\omega}{2\pi i}
<{n}|{\bf p}G(E+\omega){\bf\hat D}(\omega,k)
+{\bf\hat D}(\omega,k)G(E+\omega){\bf p}|n>
\end{equation}

and first calculate the contour integral\footnote{Note that the overall 
minus sign is connected with the respective sign in the definition of the 
transverse propagator.}

\begin{equation}            \label{tr}
\Delta E_{magn}=
-\frac{Z\alpha}{2M}<{n}|{\bf p}[\sum_+\frac{|m><m|}{k+E_m-E} 
-\sum_-\frac{|m><m|}{E-E_m+k}]\frac{4\pi\mbox{\boldmath 
$\alpha_k$}}{k}|{n}>+h.c..
\end{equation} 

As we are again calculating the low-frequency corrections to the Breit 
potential let us expand the positive energy term in $(E_m-E)/k$

\begin{equation}
\Delta E^+_{magn}
\end{equation}
\[
=-\frac{Z\alpha}{2M}<{n}|{\bf p}\sum_+|m><m|
[\frac{1}{k}-\frac{E_m-E}{k^2}+\frac{(E_m-E)^2}{k^3} 
+\ldots]\frac{4\pi\mbox{\boldmath $\alpha_k$}}{k}|{n}>+h.c..
\]

The first term in this expansion may be written in the form

\begin{equation}                 \label{breittr}
\Delta E^+_{magn1}=-\frac{Z\alpha}{2M}<{n}|{\bf p}\Lambda_+
\frac{4\pi\mbox{\boldmath $\alpha_k$}}{k^2}|{n}>+h.c.
\end{equation}
\[
=-\frac{Z\alpha}{2M}<{n}|{\bf p}\frac{4\pi
\mbox{\boldmath $\alpha_k$}}{k^2}|{n}>
+\frac{Z\alpha}{2M}<{n}|{\bf p}\Lambda_-
\frac{4\pi\mbox{\boldmath $\alpha_k$}}{k^2}|{n}>+h.c.
\]
\[
=\Delta E_{Br}+\Delta E^+_{magn1-},
\]

and it is now evident that the first (Breit) term here coincides with that 
part of transverse exchange which cancels with the respective term in the 
Coulomb contribution.

Remaining positive-energy contributions are given by the expression

\begin{equation}
\Delta E^+_{magnr}=-\frac{Z\alpha}{2M}<{n}|{\bf p}\sum_+|m><m|
[-\frac{E_m-E}{k^2}+\frac{(E_m-E)^2}{k^3} 
+\ldots]\frac{4\pi\mbox{\boldmath $\alpha_k$}}{k}|{n}>+h.c.
\end{equation}
\[
\equiv \Delta E^+_{magn2} +\Delta E^+_{magn3}+\ldots.
\]

\subsection{Positive Energy Contribution}

In accordance with Ref.\cite{elkh} one may check that the term $\Delta 
E^+_{magn2}$ does not lead to the contributions of order $(Z\alpha)^6$. 
We have 

\begin{equation}
\Delta E^+_{magn2}=\frac{Z\alpha}{2M}<{n}|{\bf p}\sum_+|m><m|
(E_m-E)\frac{4\pi\mbox{\boldmath $\alpha_k$}}{k^3}|{n}>+h.c.
\end{equation}
\[
=-\frac{(Z\alpha)^2}{2M}<{n}|\frac{4\pi\mbox{\boldmath $k'$}}{k'^2}
\Lambda_+\frac{4\pi\mbox{\boldmath $\alpha_k$}}{k^3}|{n}>+h.c..
\]

The simplest way to estimate this matrix element is to make a Fourier 
transformation. Then we need an infrared divergent Fourier transform of 
$1/k^3$.  All momentum integrals in the low-frequency region are cut off 
from below by $m(Z\alpha)^2$, and it is easy to check that the leading 
term in the infrared divergent Fourier transform generates a logarithmic 
divergent contribution of order $(Z\alpha)^5$ in accordance with 
Ref.\cite{elkh}. The next terms vanish with the infrared cutoff and cannot 
produce contributions of order $(Z\alpha)^6$.

Let us turn now to the term $\Delta E^+_{magn3}$. Naive calculation in the 
coordinate space in accordance with the result in Ref.\cite{elkh} leads to 
the result 

\begin{equation}                     \label{magn3}
\Delta E^+_{magn3}=-\frac{Z\alpha}{2M}<{n}|{\bf p}\sum_+(E_m-E)^2|m><m|
\frac{4\pi\mbox{\boldmath $\alpha_k$}}{k^4}|{n}>+h.c.
\end{equation}
\[
=-\frac{(Z\alpha)^2}{4m^2M}
<{n}|2{\bf p}\frac{1}{r^2}{\bf p}-\frac{7\bf l^2}{2r^4} 
-\frac{\mbox{\boldmath $\sigma l$}}{r^4}|{n}>. 
\]

This expression contains only operators which are nonsingular at the origin 
for $S$-states. Hence, they are well defined, and there is no need for a
careful momentum space consideration in this case.

\subsection{Negative Energy Contribution}

There are two negative-energy contributions connected with the magnetic 
term, one in Eq.(\ref{tr}), and the other in Eq.(\ref{breittr}).

Let us consider first

\begin{equation}
\Delta E^-_{magn}=\frac{Z\alpha}{2M}<{n}|{\bf p}
\sum_-\frac{|m><m|}{E-E_m+k}\frac{4\pi\mbox{\boldmath 
$\alpha_k$}}{k}|{n}>+h.c..
\end{equation}

We have checked, in accordance with Ref.\cite{elkh}, that this term leads at 
most to contributions of order $(Z\alpha)^6$, and, hence, is of no interest.

We still have to calculate one more negative energy contribution, contained 
in Eq.(\ref{breittr}) 

\begin{equation}  \label{negencontr}
\Delta E^+_{magn1-}
=\frac{Z\alpha}{2M}<{n}|{\bf p}\Lambda_-
\frac{4\pi\mbox{\boldmath $\alpha_k$}}{k^2}|{n}>+h.c.
\end{equation}
\[
=\frac{(Z\alpha)^2}{8m^2M}
<{n}|\frac{4\pi\mbox{\boldmath $\alpha_{k'}$}}{k'^2}\frac{4\pi{\bf 
k}\mbox{\boldmath $(\alpha k)$}}{k^2}|{n}>+h.c..  
\]

Naive calculation with the help of the Fourier transformation leads, in 
accordance with Ref.\cite{elkh}, to the expression

\begin{equation}               \label{naivemagn}
\Delta E^+_{magn1-}
=\frac{(Z\alpha)^2}{4m^2M}
<{n}|\frac{4\pi\delta({\bf r})}{r}-\frac{1}{r^4}|n>.
\end{equation}

However, this expression, as in the case of the seagull 
contribution, contains singular operators at the origin, and does not have 
unambiguous meaning for the $S$-states. A more careful calculation, which 
explicitly takes into account a momentum space cutoff $\sigma$, is needed.

First we transform the negative energy contribution in Eq.(\ref{negencontr}) 
to the form

\begin{equation}
\Delta E^+_{magn1-}
=-\frac{Z\alpha}{4mM}
<{n}|[{\bf p},V]\Lambda_-\frac{4\pi\mbox{\boldmath $\alpha_k$}}{k^2}|{n}>
+h.c.
\end{equation}

Next we substitute the negative energy projection operator in the 
nonrelativistic approximation $\Lambda_-({\bf p})\approx{1}/{2}
-({\mbox{\boldmath $\alpha p$}+\beta m})/{2m}$ and use the trivial identity

\begin{equation}
[p,V]\Lambda_-=\Lambda_-[p,V]-[\Lambda_-,[p,V]]=\Lambda_-[p,V]+
[\frac{\mbox{\boldmath $\alpha p$}}{2m},[p,V]].
\end{equation}

Note that the first term on the right hand side vanishes applied to the 
ket-vector, and the negative energy contribution reduces in the 
nonrelativistic approximation to 

\begin{equation}
\Delta E^+_{magn1-}
=-\frac{Z\alpha}{2m^2M}\int\frac{d^3k}{(2\pi)^3}
<{n}|{\mbox{\boldmath $p_k$}}[{\bf 
p},V]\frac{4\pi e^{i{\bf kr}}}{k^2}|{n}>.  
\end{equation}

Then we use

\begin{equation}
<n(r)|{\bf p_k}=-i\gamma<n(r)|\frac{\bf r_k}{r},
\end{equation}
\[
[{\bf p},V]=-i(Z\alpha)\frac{\bf r}{r^3},
\]

and obtain

\begin{equation}   
\Delta E^+_{magn1-}
=\frac{(Z\alpha)^2}{2m^2M}\gamma\int\frac{d^3k}{(2\pi)^3}\int d^3r
\psi(r)^2\frac{{\mbox{\boldmath ($r_kr$)}}}{r^4}\frac{4\pi e^{i{\bf 
kr}}}{k^2}.
\end{equation}

As in the case of the singular seagull contribution we will perform 
the calculation for $n=1$ first, postponing consideration of the general 
case to the next chapter. We substitute explicit expressions for the wave 
functions and obtain

\begin{equation}   \label{magnsing}
\Delta E^+_{magn1-}
=\frac{2\pi(Z\alpha)^2}{2m^2M}\gamma|\psi(0)|^2
\int\frac{d^3k}{(2\pi)^3}\frac{4\pi }{k^2} 
\int_{-1}^1dx (1-x^2)\int_0^\infty dre^{-2\gamma r}e^{ikrx} 
\end{equation}
\[
=\frac{4\pi(Z\alpha)^2}{2m^2M}\gamma^2|\psi(0)|^2
\int\frac{d^3k}{(2\pi)^3}\frac{4\pi }{k^2} 
[\frac{(4\gamma^2+k^2)\arctan\frac{k}{2\gamma}}{\gamma k^3}-\frac{2}{k^2}]
\]
\[
=\frac{4\pi(Z\alpha)^2}{2m^2M}\frac{(4\pi)^2}{(2\pi)^3}\gamma^2|\psi(0)|^2
\int_0^\sigma dk
[\frac{(4\gamma^2+k^2)\arctan\frac{k}{2\gamma}}{\gamma k^3}-\frac{2}{k^2}]
\]
\[
=\frac{\pi(Z\alpha)^2}{m^2M}\gamma|\psi(0)|^2
[2\ln\frac{\sigma}{2\gamma} -1].
\]

Again, as in the case of the seagull contribution, this term may be 
understood as a proper regularization of the naive singular in the 
coordinate space operator from Eq.(\ref{naivemagn}).

\section{Calculations for Arbitrary Principal Quantum Number}

The total low-frequency contribution for the $1S$-state is given 
by the sum of the results in Eq.(\ref{gy}), Eq.(\ref{s1}), Eq.(\ref{s3}), 
Eq.(\ref{magn3}) and Eq.(\ref{magnsing}) 

\begin{equation}   \label{totlow1s}
\Delta E_{low-freq}(1S)=
-{(Z\alpha)^6}\frac{m}{M}m,
\end{equation}

and coincides with the result obtained earlier for the low-frequency 
contribution in Ref.\cite{pg}. We see that the seagull and magnetic 
contributions partially cancel each other. This reflects cancellation of the 
$1/r^4$ terms in the language of Ref.\cite{elkh}. However,  the contribution 
$(-1)$ survives. This contribution is connected with the $\delta$-function 
term in Ref.\cite{elkh}, and the error in 
Ref.\cite{elkh} is due to an improper regularization of this contribution.  
Note that from the point of view of the coordinate representation after the 
Fourier transformation is done the proper regularization is highly 
nontrivial.  One could never obtain this contribution with a naive {\it 
ad hoc} regularization in coordinate space. 

The result in Eq.(\ref{totlow1s}) is valid only for the $1S$-state. We are 
going to generalize it to an arbitrary principal quantum number.

\subsection{Seagull Contribution for Arbitrary $nS$-Level}

The general expression for the wave function of an $nS$-level has the form

\begin{equation}
\psi_n(r)=(\frac{\gamma^3}{\pi n^3})^\frac{1}{2}e^{-\frac{\gamma r}{n}}
[1-\frac{n-1}{n}\gamma r+\ldots].
\end{equation}

Let us introduce $\beta\equiv \gamma/n$. Almost all  
calculations above for $n=1$ immediately turn into calculations for 
arbitrary $n$ after substitution $\gamma\rightarrow \beta$ \cite{ey}. The 
wave function has the form

\begin{equation}
\psi_n(r)=(\frac{\beta^3}{\pi})^\frac{1}{2}e^{-{\beta r}}
[1-(n-1)\beta r+\ldots]\equiv \psi_n(0)e^{-{\beta r}}
[1-(n-1)\beta r+\ldots].
\end{equation}

Quadratic and higher order terms in $r$ in the postexponential factor in the 
wave function do not produce any contribution to the energy level connected 
with the singular operator in the naive expression in 
Eq.(\ref{seagullop}), and we will ignore them below. The 
only difference between the general case and the case of $n=1$ is connected 
with the linear term in the postexponential factor. Let us find out how it 
changes the result for the seagull contribution. First, let us write down 
the singular seagull contribution induced by the purely exponential part of 
the wave function for arbitrary $n$ in the form

\begin{equation}
\Delta E^{1/r^4}_{s}=
-\frac{(Z\alpha)^2}{m^2M}|\psi(0)|^2
\int\frac{d^3k'}{(2\pi)^3}
\frac{d^3k}{(2\pi)^3}\frac{4\pi^2({\bf k'k})}{k'^2k^2}\int d^3r
e^{i{\bf (k+k')r}}e^{-2\beta r}
\end{equation}
\[
=-\frac{(Z\alpha)^2}{m^2M}|\psi(0)|^2\epsilon^{1/r^4}_{s},
\]

where 

\begin{equation}
\epsilon^{1/r^4}_{s}=2{\pi}\beta\ln\frac{\sigma}{2\beta} 
\end{equation}

The linear terms in the wave functions lead to an additional contribution

\begin{equation}
\Delta E^{1/r^4}_{s,corr}=
-\frac{(Z\alpha)^2}{m^2M}|\psi(0)|^2
\int\frac{d^3k'}{(2\pi)^3}
\frac{d^3k}{(2\pi)^3}\frac{4\pi^2({\bf k'k})}{k'^2k^2}\int d^3r
e^{i{\bf (k+k')r}}e^{-2\beta r}[-2(n-1)\beta r]
\end{equation}
\[
=-\frac{(Z\alpha)^2}{m^2M}|\psi(0)|^2(n-1)\beta
\frac{\partial}{\partial\beta}\epsilon^{1/r^4}_{s}
=-\frac{(Z\alpha)^2}{m^2M}|\psi(0)|^2(n-1)[2\pi\beta\ln\frac{\sigma}{2\beta}
-2\pi\beta],
\]

and the total seagull contribution to the energy shift is equal to

\begin{equation} \label{s3n}
\Delta E^{1/r^4}_{s,tot}=\Delta E^{1/r^4}_{s}
+\Delta E^{1/r^4}_{s,corr}=-\frac{(Z\alpha)^2}{m^2M}|\psi(0)|^2
[\epsilon^{1/r^4}_{s}+(n-1)\beta
\frac{\partial}{\partial\beta}\epsilon^{1/r^4}_{s}]
\end{equation}
\[
=-\frac{(Z\alpha)^2}{m^2M}|\psi(0)|^2[2\pi\gamma\ln\frac{\sigma}{2\beta}
-2\pi(n-1)\beta].
\]

\subsection{Magnetic Contribution for Arbitrary $nS$-Level}

As in the case of the seagull contribution the only difference of the 
general case from the case of $n=1$ is connected with the linear term in 
the postexponential factor in the wave function. The purely exponential part 
of the wave function leads to the following singular magnetic contribution 
for arbitrary $n$

\begin{equation}   
\Delta E^+_{magn1-}
=\frac{(Z\alpha)^2}{m^2M}|\psi(0)|^2
[2\pi\beta\ln\frac{\sigma}{2\beta} -\pi\beta]
\equiv\frac{(Z\alpha)^2}{m^2M}|\psi(0)|^2\epsilon^+_{magn1-}.
\end{equation}

The new contribution induced by the linear term in the wave function has the 
form

\begin{equation}
\Delta E^+_{magn1-,cor}
=-\frac{Z\alpha}{2m^2M}\int\frac{d^3k}{(2\pi)^3}
[-(n-1)\beta]\{<{n}|r{\mbox{\boldmath $p_k$}}[{\bf 
p},V]\frac{4\pi e^{i{\bf kr}}}{k^2}|{n}>
\end{equation}
\[
+<{n}|{\mbox{\boldmath $p_k$}}[{\bf 
p},V]\frac{4\pi e^{i{\bf kr}}}{k^2}r|{n}>\}.  
\]

Next we write

\begin{equation}
r{\bf p_k}={\bf p_k}r-[{\bf p_k},r],
\end{equation}

and using the commutation relation

\begin{equation}
[{\bf p_k},r]=-i\frac{\bf r_k}{r},
\end{equation}

obtain

\begin{equation}
\Delta E^+_{magn1-,cor}
=-\frac{Z\alpha}{2m^2M}\int\frac{d^3k}{(2\pi)^3}
[-(n-1)\beta]\{<{n}|i\frac{\bf r_k}{r}[{\bf 
p},V]\frac{4\pi e^{i{\bf kr}}}{k^2}|{n}>
\end{equation}
\[
+<{n}|{\mbox{\boldmath $p_k$}}[{\bf 
p},V]\frac{4\pi e^{i{\bf kr}}}{k^2}2r|{n}>\}.  
\]
\[
=-\frac{Z\alpha}{2m^2M}\int\frac{d^3k}{(2\pi)^3}
(n-1)\{<{n}|(-i\beta\frac{\bf r_k}{r})[{\bf 
p},V]\frac{4\pi e^{i{\bf kr}}}{k^2}|{n}>
\]
\[
-(n-1)\beta<{n}|{\mbox{\boldmath $p_k$}}[{\bf 
p},V]\frac{4\pi e^{i{\bf kr}}}{k^2}2r|{n}>\}  
\]
\[
=\frac{(Z\alpha)^2}{m^2M}|\psi(0)|^2
(n-1)[\epsilon^+_{magn1-}+\beta^2\frac{\partial}{\partial\beta}
(\frac{\epsilon^+_{magn1-}}{\beta})]
\]
\[
=\frac{(Z\alpha)^2}{m^2M}|\psi(0)|^2
[2\pi\beta(n-1)\ln\frac{\sigma}{2\beta} 
-(n-1)\pi\beta-(n-1)2\pi\beta]
\]
\[
=\frac{(Z\alpha)^2}{m^2M}|\psi(0)|^2
[2\pi\beta(n-1)\ln\frac{\sigma}{2\beta} 
-3(n-1)\pi\beta].
\]

Then the total singular magnetic contribution is equal to

\begin{equation}   \label{magnsingn}
\Delta E^+_{magn1-,tot}=\Delta E^+_{magn1-}+\Delta E^+_{magn1-,cor}
=\frac{(Z\alpha)^2}{m^2M}|\psi(0)|^2
[2\pi\beta\ln\frac{\sigma}{2\beta} -\pi\beta
\end{equation}
\[
+2\pi\beta(n-1)\ln\frac{\sigma}{2\beta} -3(n-1)\pi\beta]
=\frac{(Z\alpha)^2}{m^2M}|\psi(0)|^2
[2\pi\gamma\ln\frac{\sigma}{2\beta} -\pi\beta-3(n-1)\pi\beta].
\]

\section{Total Recoil Correction }

The total low-frequency contribution of order $(Z\alpha)^6(m/M)m$ for 
arbitrary $nS$-state is given by the sum of the terms in Eq.(\ref{gy}), 
Eq.(\ref{s1}), Eq.(\ref{s3n}), Eq.(\ref{magn3}) and Eq.(\ref{magnsingn}) 

\begin{equation}   \label{totlow}
\Delta E_{low-freq}=
(\frac{1}{8}+\frac{3}{8n}-\frac{1}{n^2}+\frac{1}{2n^3})
\frac{(Z\alpha)^6}{n^3}\frac{m}{M}m
+\frac{(Z\alpha)^2}{2m^2M}<n|{\bf p}\frac{1}{r^2}{\bf p}|n>
\end{equation}
\[
-\frac{(Z\alpha)^2}{m^2M}|\psi(0)|^2[2\pi\gamma\ln\frac{\sigma}{2\beta}
-2\pi(n-1)\beta-2\sigma]
-\frac{(Z\alpha)^2}{4m^2M}
<{n}|2{\bf p}\frac{1}{r^2}{\bf p}|{n}>
\]
\[
+\frac{(Z\alpha)^2}{m^2M}|\psi(0)|^2
[2\pi\gamma\ln\frac{\sigma}{2\beta} -\pi\beta-3(n-1)\pi\beta]
\]
\[
=(\frac{1}{8}+\frac{3}{8n}-\frac{1}{n^2}+\frac{1}{2n^3})
\frac{(Z\alpha)^6}{n^3}\frac{m}{M}m-\frac{(Z\alpha)^6}{n^3}\frac{m}{M}m.
\]

Note that the last term connected with the naive singular operators in the 
coordinate space turned out to be state-independent.

To obtain the total recoil correction of order $(Z\alpha)^6(m/M)m$ 
it is also necessary to calculate the high-frequency (or short-distance) 
contribution to the energy shift. The simplest way is to use again the Braun 
formula Eq.(\ref{braun}), but this time in the Feynman gauge. This 
calculation is quite straightforward if one again uses the auxiliary 
parameter $\sigma$ introduced above in order to qualify would be infrared 
divergences. Such a calculation was performed explicitly in ref.\cite{elkh} 
and led to the result

\begin{equation}                 \label{tothigh}
\Delta E_{high-freq}=(4\ln2-\frac{5}{2})\frac{(Z\alpha)^6}{n^3}\frac{m}{M}m,
\end{equation}

in complete agreement with Ref.\cite{pg}. 

Then total correction of order $(Z\alpha)^6(m/M)m$ to the energy levels is 
given by the sum of the results in Eq.(\ref{totlow}) and Eq.(\ref{tothigh})

\begin{equation}                 \label{tot}
\Delta E_{tot}=(\frac{1}{8}+\frac{3}{8n}-\frac{1}{n^2}+\frac{1}{2n^3})
\frac{(Z\alpha)^6}{n^3}\frac{m}{M}m
+(4\ln2-\frac{7}{2})\frac{(Z\alpha)^6}{n^3}\frac{m}{M}m.
\end{equation}

For $n=1,2$ this result nicely coincides with the one obtained in 
Ref.\cite{pg}. 

In conclusion let us emphasize that discrepancies between the different 
results for the correction of order $(Z\alpha)^6(m/M)$ to the energy levels 
of the hydrogenlike ions are resolved and the correction of this order is 
now firmly established. 

\acknowledgements
M. E. is deeply grateful for the kind hospitality of the Physics 
Department at Penn State University, where this work was performed. The 
authors appreciate the support of this work by the National Science 
Foundation under grant number PHY-9421408.

\end{document}